\documentclass[10pt,journal,compsoc]{IEEEtran}

\usepackage{graphicx}
\usepackage{amsmath,amssymb}

\usepackage{tabularx}
\usepackage{multirow}
\usepackage{dcolumn}
\usepackage{booktabs}
\usepackage{enumerate}

\usepackage{subfig}
\graphicspath{{./images/}}

\usepackage{pgfplots}
\pgfplotsset{compat=1.12}
\usetikzlibrary{patterns}
\usetikzlibrary{spy}

\usepackage{tikz}

\usepackage{algpseudocode}
\usepackage{algorithm}

\algnewcommand\algorithmicforeach{\textbf{for each}}
\algdef{S}[FOR]{ForEach}[1]{\algorithmicforeach\ #1\ \algorithmicdo}

\usepackage{amsthm}

\usepackage{nomencl}
\makenomenclature

\usepackage{etoolbox}
\renewcommand\nomgroup[1]{%
  \item[\bfseries
  \ifstrequal{#1}{A}{Acronyms}{%
  \ifstrequal{#1}{S}{Symbols}{%
  \ifstrequal{#1}{U}{Subscripts and Superscripts}{}}}%
]}

\usepackage{hyperref}
\hypersetup{
  urlcolor   = blue,
  colorlinks = true,
}

\usepackage[normalem]{ulem}

\ifCLASSOPTIONcompsoc
  \usepackage[nocompress]{cite}
\else
  \usepackage{cite}
\fi

\begin{document}

\title{Instanced model simplification using combined geometric and appearance-related metric}

\author{
Sadia Tariq, Anis Ur Rahman, Tahir Azim and Rehman Gull Khan \\

\thanks{Sadia Tariq and Anis U. Rahman are with School of Electrical Engineering and Computer Science (SEECS), National University of Sciences and Technology (NUST), Islamabad, Pakistan. Tahir Azim is with \'{E}cole Polytechnique F\'{e}d\'{e}rale de Lausanne (EPFL), Lausanne, Switzerland. Rehman G. Khan is with Bahauddin Zakaria University (BZU), Multan, Pakistan

Corresponding Author: Anis Ur Rahman, e-mail: anis.rahman@seecs.edu.pk}
%
}
\maketitle

\abstract{
Evolution of 3D graphics and graphical worlds has brought issues like content optimization, real-time processing, rendering, and shared storage limitation under consideration. Generally, different simplification approaches are used to make 3D meshes viable for rendering. However, many of these approaches ignore vertex attributes for instanced 3D meshes. In this paper, we implement and evaluate a simple and improved version to simplify instanced 3D textured models. The approach uses different vertex attributes in addition to geometry to simplify mesh instances. The resulting simplified models demonstrate efficient time-space requirements and better visual quality.
}


\begin{IEEEkeywords}
3D meshes, simplification, instancing
\end{IEEEkeywords}

\IEEEpeerreviewmaketitle

\section{Introduction}
\label{sec:intro}

\noindent Technology brings imagination closer to reality. This phenomenon can be demonstrated through the presentation of 3D models. Traditionally, the use of virtual worlds for social communication is considered in the context of 3D games. However, these computer-based graphical environments also provide other morphs of interaction to participants, including forums, blogs, wikis, chat rooms, instant messaging, and video conferencing. Such platforms help build communities with like-minded people who can share information and gain new experiences. Likewise, from advertising and marketing to geological hazard simulations, information is presented as 3D data that must be shared in order to bring the imagined project to reality. In particular, the application of a virtual environment is most obvious in the entertainment and gaming world. However, latency in transferring 3D data is a major issue in this industry. Moreover, most content of virtual worlds is created by many users requiring shared data storage. The visual quality of the 3D models suffers as a result of latency in such shared data storages.

Computing large amounts of client-produced cloud-based graphical data highlights the challenges in storage, network, and processing; that is, user-developed models are not optimized for rendering in terms of face count. For instance, virtual worlds contain a large number of 3D meshes/objects in a scene where many of these are repeated with the same geometry but different sizes, orientations, and other properties. Such repeated meshes are referred to as instanced 3D meshes. Modern 3D tools take a mesh with multiple copies as instanced input and save this data as one mesh with different properties. This approach saves memory and renders livelier scenes. However, a major challenge faced while simplifying instanced 3D meshes is dealing with a huge number of copies as a single mesh. Another challenge relates to sorting out the meshes into groups; that is, deciding which object is an exact copy of which one. Subsequently, downloading just a few of them can overburden the network bandwidth and graphic processors. Thus, transferring huge amounts of data with detailed textures and normal attributes is not feasible.

Numerous attempts with optimized results have previously been made in the field of 3D model simplification. As aforementioned, the most common and serious issue with many of these approaches is the resulting file size. In some cases, the file size doubles, in particular when mesh instancing is not taken into account. Present techniques for reducing meshes and making them appropriate for the cloud presents different challenges. These techniques efficiently reduce the number of triangles from the mesh to produce good visual results, but many of these techniques increase the file size--making it unfit for shared data storage~\cite{erikson99,garland97,ronfard96}. This is not a concern when the data is stored on local disks. However, with the advent of virtual worlds and real-time multi-dimensional environments, the data needs to be shared to ensure robustness and to maintain the integrity of the user-developed content. This requirement makes cloud new storage points, altering the direction of research from simple simplification to efficient simplification with reduced data sizes to avoid latencies. To resolve this issue, the approach in~\cite{azim15} considers instancing and coherence in real-world to decrease file size by reducing the triangle count of meshes. This 3D simplification, however, does not take into account vertex attributes such as textures and colors. There is a need to develop techniques that also consider these attributes while simplifying 3D content to produce smaller sized results.

In this paper, we propose an instanced simplification approach that incorporates both instancing and the concept of simplifying different vertex attributes. The former records copies of the same mesh as references rather than as complete data. This provides an opportunity to have faster processing times by simplifying multiple copies of the same mesh and to save storage space by keeping only references of multiple clones of the mesh. While the latter simplifies an instanced mesh by reducing its triangle count to get smaller files, as well as, addresses different attributes such as the texture, color, and normals.

The main contribution of this work is a simple and efficient algorithm that accounts for vertex attributes when simplifying instanced 3D textured models. A customized format was used to work with readable instanced input and output for an accurate understanding of the variations which occur during the simplification procedure. In summary, the proposed technique uses instances, vertex attributes, and geometrical simplification while producing smaller sized result files.

The remainder of the paper is organized as follows: in Section~\ref{sec:related}, we provide the theoretical foundations and a review of the state-of-the-art algorithms for 3D mesh simplification. Our main contributions are presented in Section~\ref{sec:proposed}. In Section~\ref{sec:results}, we report comprehensive experimental results to validate the proposed approach, followed by discussion in Section~\ref{sec:discussion}. Finally, in Section~\ref{sec:conclusion}, conclusions and future research directions are outlined.

\section{Related work}
\label{sec:related}


Since the evolution of multi-dimensional presentation and the ways of communication over the last decade, when live 3D environments and virtual worlds have taken initiatives, 3D simplification algorithms have been well-studied. However, increased usage of this concept has come with new limitations, opening an entirely untouched dimension for research. 

A lot of previous research has been done on geometrical simplification of 3D models. This includes simplification of vertex attributes and other local properties~\cite{erikson99,garland97}. But, the major issues with these methods came to light when data was transferred over the network for sharing. The methods either double the resultant file sizes or ignore different attributes of vertices while simplifying 3D meshes. However, in~\cite{azim15}, file sizes after simplification of 3D models are efficiently handled by instance-aware simplification where common factors of coherence and instancing of meshes in a large 3D world are exploited. On the downside, without simplification of attributes, a significant amount gets added in data size. 

We can organize 3D simplifications into three major categories: a)~geometric simplification, b)~attribute simplification, and c)~instanced mesh simplification. 

\subsection{Geometric Simplification}

Traditionally, research in 3D model simplification involves geometric data reduction with predefined conditions and assumptions. Schroeder et al.~\cite{schroeder92} proposed one of the first iterative decimation approach using vertex-to-plane distances. This approach is fast, producing vertices as a subset of the original set of vertices, but changes to the vertex positions limit the fidelity of simplification. Generally, selection methods for geometric simplification work by producing an optimal point for collapse of a vertex pair. The methods are efficient at producing quality results by classifying vertices into a hierarchal triangulation. All vertices contained are decimated iteratively followed by re-triangulation of the remaining vertices to fill any holes. Other selection approaches use different metrics to guide this decimation: curvature-based~\cite{reddy96}, Hausdorff distance-based~\cite{klein96}, local tessellation and geometric error~\cite{ronfard96}, volume~\cite{gueziec96,lindstrom1998fast}, and distances to supporting forms~\cite{rossignac97}. 

In~\cite{pajarola00}, decimation is implemented using group vertex splits encoded as vertex spanning trees. A similar approach~\cite{taubin98a} using vector quantization, instead of scalar quantization used earlier, results in faster decoding and lower memory usage but lower simplification fidelity. Another extension introduces progressive vertex forest splits to get higher compression rates~\cite{taubin98b}. Though the fidelity improves but the process is slower; however, there exists an efficient implementation exploiting hardware vertex buffers to improve vertex data reuse~\cite{karni02}. In~\cite{li06}, a more sophisticated data-driven approach is proposed using vector quantization with arbitrarily shaped groups for encoding. This results in better rate-distortion performance but the algorithm is computationally expensive.

More recent works in simplification literature use multi-resolution schemes~\cite{valette04,kim11} and improved quantization schemes~\cite{lee11,lee12,ahn10} for intermediate meshes to reduce distortion. Other improvements include the use of Gaussian mixture models for vertex creation~\cite{Lee14}, patch coloring for iterative decimation~\cite{cohen99} followed by cleansing conquest scheme in~\cite{alliez01} and null patch avoidance scheme proposed in~\cite{ahn11}. Many of these methods improve bit-rates but the geometry is somewhat degraded. This is countered by the use of simplification envelopes~\cite{cohen96,bahirat2017boundary} to guide the simplification process for irregular meshes, ensuring fidelity bounds for boundary and texture preservation.

The downside to the aforementioned geometric simplification algorithms is that they are limited to manifold surfaces; that is, all such methods can close gaps created after vertex removal using successive edge contraction but are unable to join unconnected faces. Furthermore, many of them fail to meet all model parameters defined in~\cite{campomanes2013evolutionary} including quality, efficiency, generality, and optimization of geometric simplification.

\subsection{Attribute Simplification}

Classical quadric error method~\cite{garland97} is efficient with reasonably good results, later followed by an extension incorporating different appearance attributes~\cite{garland98}. Similarly, Hoppe~\cite{hoppe96} introduces progressive meshes for storage and transmission of triangle meshes. The approach uses successive edge collapses based on energy functions defined on these attributes to preserve granularity. But, the trivial decimation process used becomes expensive when the number of candidate vertices for collapse approach the total model size. 

An adaptive threshold selection scheme proposed in~\cite{erikson99} uses a surface-area metric complemented with other shading attributes to improve simplification fidelity. Other similar methods introduce appearance preserving simplifications~\cite{garland98,lindstrom00,sander01}. The process may lead to loss of geometric details, degrading actual appearance. This issue was alleviated in a generalized quadric approach~\cite{hoppe99} based on appearance attributes to accurately and efficiently simplify meshes while preserving fidelity.

In~\cite{cohen98}, the approach guarantees fidelity on the original appearance based on envelopes proposed in~\cite{cohen96}. These envelopes are defined using a texture stretch and deviation metric~\cite{sander01} to preserve appearance but at the cost of expensive parameterizations. The parameterizations can be avoided using different optimizations: sampling the normal map using ray casting concepts~\cite{sander00,sander01,tarini03}, or randomly sampling the normal map~\cite{cook07}, or using an adaptive sampling of a high-resolution normal map encoded as an octree~\cite{lacoste07}. The latter supports efficient lookups while avoiding expensive parameterizations. Furthermore, texture-based simplification approach~\cite{toubin98,toubin00,mao2013generalization} to visualize textures dynamically at different scales uses wavelet transformation to compress the textures and then recolor them. A surface reconstruction method proposed in~\cite{turner2015fast} produces dense meshes with preserved detail and sharpness. The approach is scalable for indoor 3D models with textures. All these methods attempt to preserve fidelity to the original appearance. 
   
\subsection{Instanced Meshes} 

The above mentioned simplification approaches take a model or a 3D environment as an indexed triangle mesh comprising a list of position points called vertices, details of textures and normal coordinates, and a list of indexed triangles. In contrast, real-world models use formats that describe a 3D model as an instanced mesh, for instance, COLLADA~\cite{arnaud06}, Maya, Autodesk 3ds Max file formats, and so forth, where the mesh structure consists of a list of sub-meshes and their corresponding transformations for instancing.

Current mesh simplification algorithms including quadric approach and GAPS are unaware of instances in their input. They generally require expansion of instancing as indexed triangles for simplification, in turn, the simplified file output requires much more storage space. Such large files result in latency during network transmission, making them unsuitable for richly instanced virtual worlds.

More recent work referred to as instance aware simplification (IAS) in~\cite{azim15} exploits the similarity of duplicated meshes and coherence among them for instance-aware simplification. To be more precise, it is an instanced variant of quadric simplification that simplifies a sub-mesh once and applies this change to all the instances referred to this sub-mesh. The resulting simplification is irreversible, reducing processing times while dealing with instanced meshes. It is an improved approach with respect to processing times and end file sizes but lacks attribute simplification.

As pointed out, none of the methods take attributes into account. To resolve this issue, we propose an improved version of GAPS that incorporates the concept of instancing to the existing approach. GAPS itself is an attribute version of quadric simplification where the attribute error cost is added into vertex geometric cost matrices. We introduce instancing to this concept to get complete simplification of all duplicated copies with no additional processing time required.

\section{Proposed Solution}
\label{sec:proposed}


We propose an extended version of GAPS as instanced textured 3D model simplification (ITS). This approach starts off with simplifying mesh geometry using the classical quadric approach followed by simplifying mesh attributes derived from GAPS, while maintaining compression using instancing. The proposed algorithm mainly comprises three major steps as follows:

\begin{enumerate}
\item \textbf{Step 1.} Simplify mesh geometry using quadric error matrices. This geometrical simplification requires some prerequisites including threshold calculation and the listing of valid pairs for merging into destination vertices.
\item \textbf{Step 2.} Calculate attribute errors via point clouds leading to a unified error. This error is the basis of merging. 
\item \textbf{Step 3.} Apply simplification across all instances using corresponding transformation matrices. 
\end{enumerate}

\begin{figure*}[!ht]
\centering
\includegraphics[width=1\textwidth]{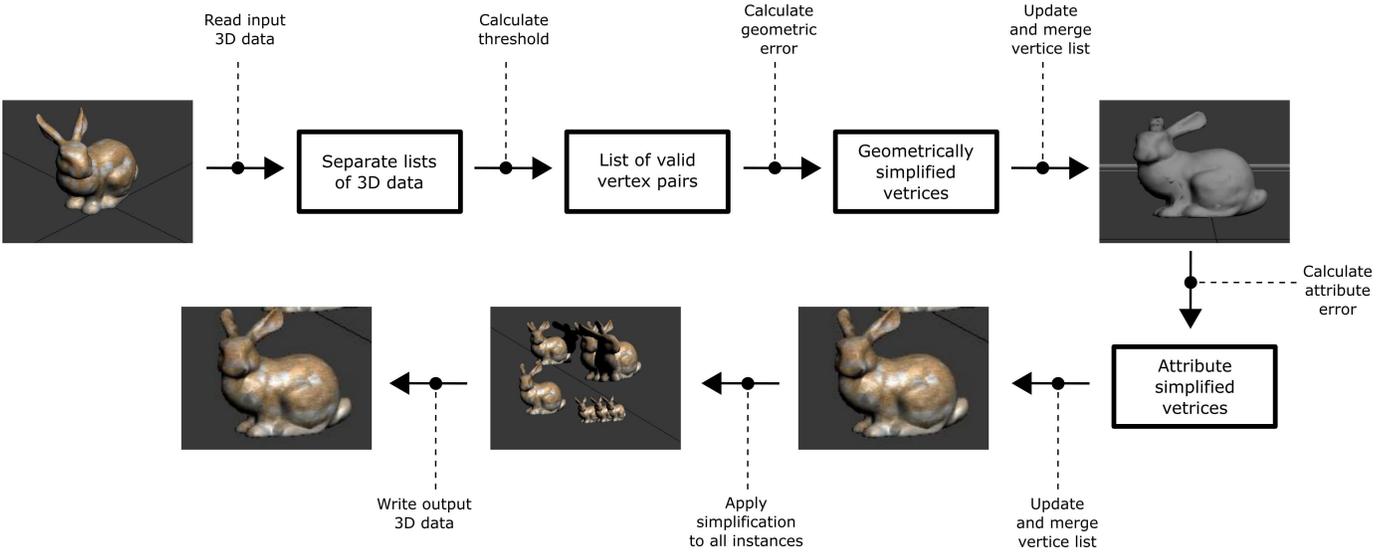}
\caption{Overview of ITS (instanced textured 3D model simplification)}
\label{fig:flow}
\end{figure*}

The basic work flow of the proposed ITS algorithm is shown in Figure~\ref{fig:flow} comprising four major phases illustrated in Algorithm~\ref{alg:proposed} including taking and interpreting input and writing details of an output file.

\algdef{SE}[SUBALG]{Indent}{EndIndent}{}{\algorithmicend\ }
\algtext*{Indent}
\algtext*{EndIndent}
\begin{algorithm}
\footnotesize
\caption{Instance Aware GAPS}
\begin{algorithmic}	
\State 1. Read Input
	\Indent
	\newline
	\State (a) Store data (vertices, textures, normal, transformations, triangles, number of instances) separately.
	\EndIndent
\newline
\State 2. Simplify the instanced mesh
	\Indent
	\newline
	\State i. Compute threshold \newline
	\State ii. Find valid pairs for simplification \newline
	\State iii. Recompute threshold if valid pairs $>10$ or $<0$ \newline
	\State iv. Compute geometric error for each vertex in valid pairs list \newline
	\State v. Compute attribute errors  (textures and normal) \newline
	\State vi. Compute unified error for each vertex \newline
	\State vii. Collapse edges with minimal error vertices \newline
	\State viii. Repeat until desired level of simplification reached within error limits
	\EndIndent
\newline
\State 3. Apply simplification to all instances \newline
\State 4. Generate output as an indexed triangles 
\end{algorithmic}
\label{alg:proposed}
\end{algorithm}

\begin{enumerate}

\item \textbf{Input/Output Format.} Often multidimensional virtual world environments take input as instanced meshes. Object file format, with a \emph{.obj} extension, is the simplest readable format to store 3D data. We customized the input object file format by adding information related to all its instances including the number of copies of a particular mesh in a given scene, their orientations, sizes and rotation angles etc. This information was stored as a matrix included in the object file format, later used to retrieve instanced 3D sub-meshes as indexed 3D files. Each sub-mesh is a result of multiplying the instance vertices information--a transformation matrix--to a given reference mesh. We produced two types of outputs: one an instanced 3D textured output and another an indexed 3D textured output carrying instance mesh information as a complete 3D mesh. The former output with extension .obj was a classical object file excluding support for instances, used to get reduced file sizes. While the latter output file, an instanced 3D textured output, was used to visualize the instances retrieved as indexed meshes. Additionally, a material file (\emph{.mtl}) was stored containing all the attribute information including textures, normals, and colors. 

\item \textbf{Errors Calculation.} The proposed algorithm applies quadric simplification followed by calculation of geometrical error of the 3D mesh. For texture and normal error, it uses point clouds. Afterward, all the computed errors are unified to select and merge the least errored vertex pairs.

\item \textbf{Instancing.} Once the sub-mesh is simplified using its attributes to a desired level, simplification is applied to all the referred instances using transformation matrices.

\end{enumerate}

\subsection{Instanced 3D Model Simplification}

The motivation of this paper is to produce an optimal simplification algorithm for textured 3D instanced meshes that minimize space requirement of the simplified output file over the cloud. It aims to introduce an approach that not only simplifies the 3D structures but also deals with vertex properties minimizing loss of information, resulting in better visual quality.

As our approach is an enhanced and instanced version of GAPS, we start with an introduction to quadric matrices for geometry simplification, then give an explanation of attributes handling and instancing, and finally conclude with a detailed work flow. 

\subsection{Quadric Simplification}

Classic quadric simplification~\cite{garland97} works in two steps: First, calculates error matrices for each vertex, and second, contracts the edge with least error.

\begin{enumerate}

\item \textbf{Quadric Error.} For each collapse, an error cost is defined, which is calculated by assigning a $4 \times 4$ matrix representing a cost against every vertex. This matrix is known as the $Q$ matrix. Error at each vertex $\mathbf{v} = [ v_x ~~ v_y ~~ v_z ~~ 1 ]^T$ in quadratic form is given as follows,
\begin{align*}
\Delta (v) &= v^T( \sum_{p \epsilon planes(v) } \textbf{K}_p) v
\end{align*}
where $\textbf{K}p$ is the matrix:
\begin{align*}
\textbf{K}_p = pp^T =\left[\begin{array}{cccc}
a^2 & ab & ac & ad	\\
ab & b^2 & bc & bd	\\
ac & bc & c^2 & cd	\\
ad & bd & cd & d^2
\end{array}\right]
\end{align*}
This fundamental error quadric $K_p$ can be used to find the squared distance of any point in the space to the plane $p$. These quadrics can be summed up to represent an entire set of planes with a single matrix $Q$.

\item \textbf{Pair Selection.} GAPS defines some rules to select valid pairs for collapse. Let us assume ($v_1$,$v_2$) is a valid pair if either ($v_1$,$v_2$) is an edge, or ${\|}v_1-v_2 {\|} < t$ where $t$ is a threshold. A vertex pair is a valid pair if and only if there exists an edge and the distance between both vertices is less than a defined threshold.

\item \textbf{Distance Threshold.} Automated and adaptive nature of threshold selection makes the simplification procedure efficient and preserves the surface area as well~\cite{erikson99}. We use similar threshold selection in this work. The threshold remains if there exists any valid pair, otherwise it gets doubled. On the other hand, if there exist more than ten valid pairs under this threshold, then the threshold gets halved.

\item \textbf{Edge Contraction.} Valid pair of vertices degenerate to one of the vertices involved in the collapse, one with a smaller error. 

\end{enumerate}

\subsection{Attributes Simplification}
Vertex attributes including their position, texture coordinates, colors, and normals are dealt in GAPS by Erikson and Manocha~\cite{erikson99}. They approached these properties by introducing the concept of point clouds. The concept is very simple and efficient, but only an approximate method for computing error in attribute space. We use a similar concept to GAPS point cloud approach for local attribute simplification.

\begin{enumerate}

\item \textbf{Locating merged vertex.} Quadric error was used to locate the position of the merged vertex. We did not allocate a new optimal point to get the pair collapsed; rather, we defined one of the vertices from the selected pair with the least error to be the destination point.

\item \textbf{Attribute error via point clouds.} A simple and efficient but approximate approach to find the attribute error is introduced by GAPS~\cite{erikson99}. The approach uses a point cloud, a pool of 3D points with some weight.\\

The approximate average error $A(p)$ at point $p$ with respect to cloud $X$ is,
\begin{align*}
A(p) = \sqrt{\frac {\Pi(p)}{X_0}}
\end{align*}
where $\Pi(p)$ is the weighted sum of squared distances of a point $p$ from the point cloud $X$. Here, $X$ at vertex $v$ is $[ X_t ~~ X_c ~~ X_n ]$ texture coordinate point cloud, color point cloud, and normal point cloud respectively. While $X_0$ is the sum of weights of all points included in the point cloud $X$. For instance, $X_{n_0}$ is the sum of normal point cloud $X_n$ and $X_{t_0}$ for texture point cloud $X_t$.

\item \textbf{Unified error metric.} At vertex $v$ with geometric error, texture coordinate point cloud $X_t$, color point cloud $X_c$, and normal point cloud $X_n$, a unified error metric defines the complete error for the collapse~\cite{erikson99}. It is a weighted average error for vertex $v$ given as:
\begin{align*}
E(v) &= \frac {S(v)\cdotp \Gamma(v) + X_{n_0}\cdotp N(v) + X_{c_0}\cdotp C(v) + X_{t_0}\cdotp T(v)}{S(v) + X_{n_0} + X_{c_0} + X_{t_0}} \\
&where \\
&S(v): \mbox{total surface area of vertex } v \\
&\Gamma(v): \mbox{geometric error of vertex } v \\
&N(v): \mbox{normal error of vertex } v \\
&T(v): \mbox{texture coordinate error of vertex } v \\
&C(v): \mbox{color error of vertex } v
\end{align*}

\item \textbf{Interpolating attributes.} To determine the new attribute value for a merged vertex, the merged vertex was projected onto its local geometry assuming the geometry prior to the merge. Limiting the search to this local geometry, we interpolated the original attributes to obtain attribute values for the new vertex. In case of attribute discontinuities at a merged vertex, the process was similar but more than one pair of attributes were merged.

\end{enumerate}

\subsection{Instancing}
This paper follows in the footsteps of IAS to deal with instanced meshes, except that it applies attribute simplification in addition to geometric simplification.

\subsection{GAPS over Instanced 3D meshes}

We achieved attribute simplification with reduced file sizes by applying concepts similar to GAPS for instance 3D meshes. Once valid pairs to collapse are nominated and their calculated geometric, texture, and normal errors are unified, the pairs with least unified error are selected for collapse. Note that it becomes impossible to revert back to the original mesh because the change is applied to all referenced instances of a particular sub-mesh by multiplying each instance with respective transformation matrices.

Suppose a scene with a sub-mesh $M$ has $n$ number of instances $\{m_1,m_2,\cdots,m_n\}$ with transformations $\{t_1,t_2,\cdots,t_n\}$. Sub-mesh $M$ has $i$ number of vertices, $j$ number of texture coordinates $\{T_1,T_2,\cdots,T_j\}$, $k$ number of normal coordinates $\{N_1,N_2,\cdots,N_k\}$, and $l$ number of faces $\{F_1,F_2,\cdots,F_l\}$. A $4 \times 4$ transformation matrix contains information about rotation, scaling, and translation of an instance. After applying simplification, sub-mesh $M^\prime$ is reduced to $i^\prime$ number of vertices, $j^\prime$ number of texture coordinates, $k^\prime$ number of normal coordinates, and $l^\prime$ number of faces. The simplified textured instanced 3D meshes are calculated using,
\begin{align*}
M^\prime \times t_i &= m_i^{t_i}
\end{align*}
The normal and texture coordinates for a particular instance may record after position coordinates. Simplified texture coordinates and normal coordinates are obtained as a result of reference submesh attribute simplification. Instances may have a different scale, rotation, and translation but they must be an exact copy of reference submesh in geometry. For simplicity, we assumed that all instances of a submesh have same texture coordinates as those of the reference submesh.

Let $T$ be the texture coordinates and $N$ be the normal coordinates, then simplification is applied to instances $I_n$ in the following routine,
\begin{align*}
&\left[\begin{array}{cccc}
M^\prime \times t_1 = m_1^{t_1} & T_1 & N_1 & F_1\\
M^\prime \times t_2 = m_2^{t_2} & T_2 & N_2 & F_2\\
M^\prime \times t_3 = m_3^{t_3} & T_3 & N_3 & F_3\\
\vdots & \vdots & \vdots & \vdots \\
M^\prime \times t_n = m_n^{t_n} & T_n & N_n & F_n
\end{array}\right]\\
\mbox{where} &\\
M^\prime &: \mbox{ simplified geometric mesh} \\
t_i &: i^{th} \mbox{ instance's transformation} \\
m_i^{t_i} &: i^{th} \mbox{ simplified instance} \\
T_i &: i^{th} \mbox{ simplified instance's textures} \\
N_i &: i^{th} \mbox{ simplified instance's normals} \\
F_i &: i^{th} \mbox{ simplified instance's face list}
\end{align*}
The resulting simplification algorithm is experimentally evaluated using input 3D models in the following section.

\section{Experimental Results}
\label{sec:results}


This section describes both qualitative and quantitative results obtained after evaluating the proposed simplification method using different inputs. The algorithm was developed in Visual Studio using VB.NET and regex was used to retrieve data. All results were compiled on a Windows 10 64-bit system with Intel i3 processor and 4GB system memory, while 3ds Max Studio and MeshLab aided visualization of the inputs and results.

We compare the results to two previous techniques: Quadric and GAPS simplification, at 10\%, 20\% and 50\% level of simplification. For the time being, the proposed system is designed only for triangulated inputs; that is, other polygons are not treated yet. Test 3D input models are shown in Table~\ref{tbl:stats} alongside their respective statistics.

\begin{table*}[!ht]%
\caption{Input 3D models and their respective statistics. \label{tbl:stats}} 
\centering
\begin{tabular*}{0.8\linewidth}{@{\extracolsep\fill}ccccccccccclD{.}{.}{3}l@{\extracolsep\fill}}
\toprule
Models	& Vertices 	& Texture 		& Normal 		& Triangles 	& \# Instances 	& Size (KB) per instance \\ [0.5ex]
\midrule
\raisebox{-.5\height}{\includegraphics[width=0.15\textwidth,height=15mm]{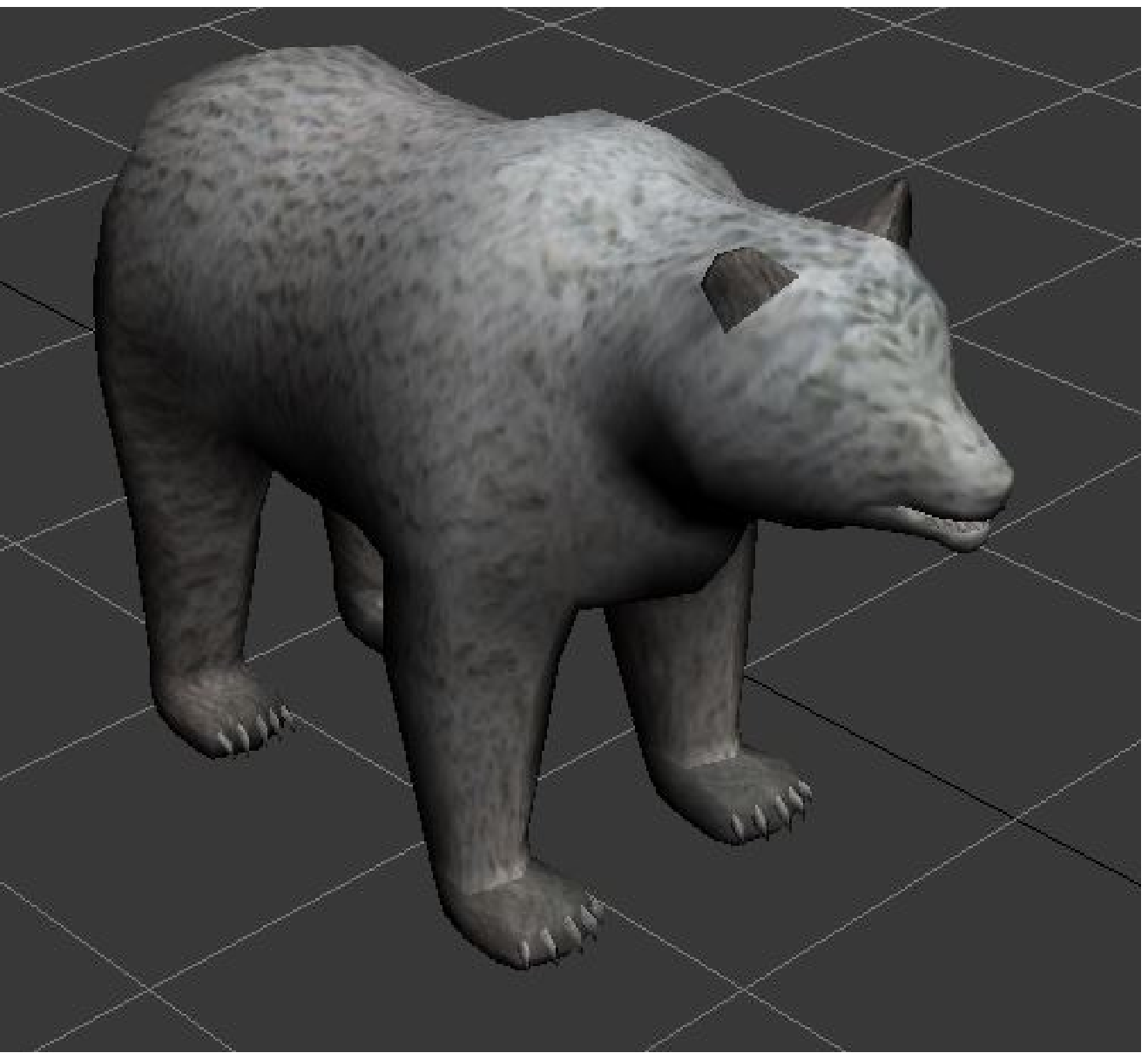}} 	&  742 &  477 &  742 & 1360 &  5 & 127 \\ \addlinespace[0.1cm]
\raisebox{-.5\height}{\includegraphics[width=0.15\textwidth,height=15mm]{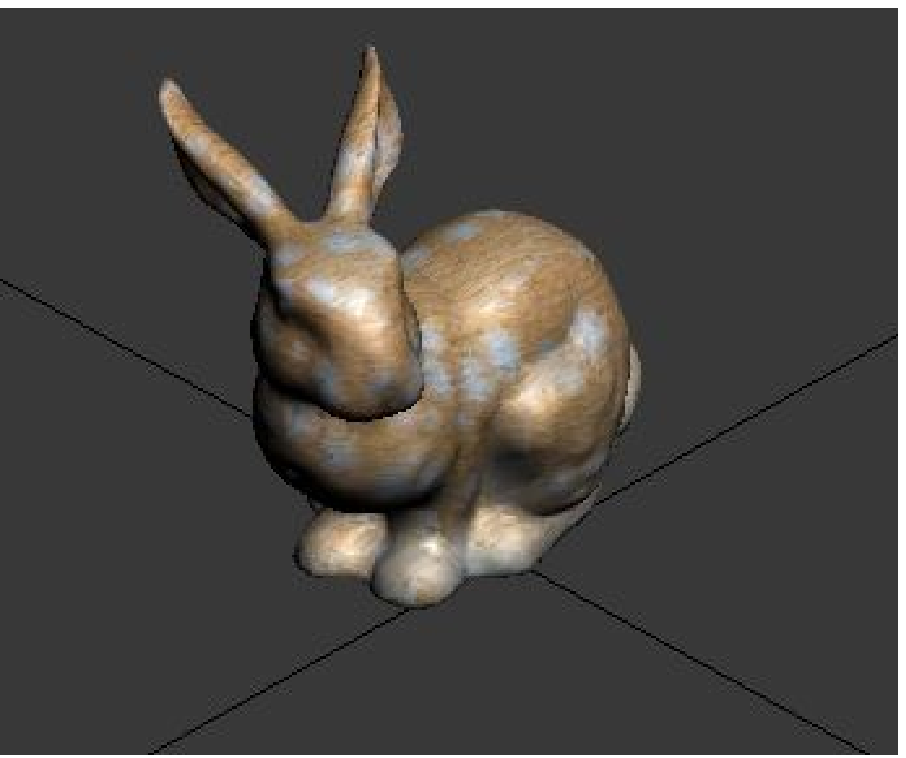}}	& 2503 & 3477 & 2503 & 4968 & 13 & 529 \\
\addlinespace[0.1cm]
\raisebox{-.5\height}{\includegraphics[width=0.15\textwidth,height=15mm]{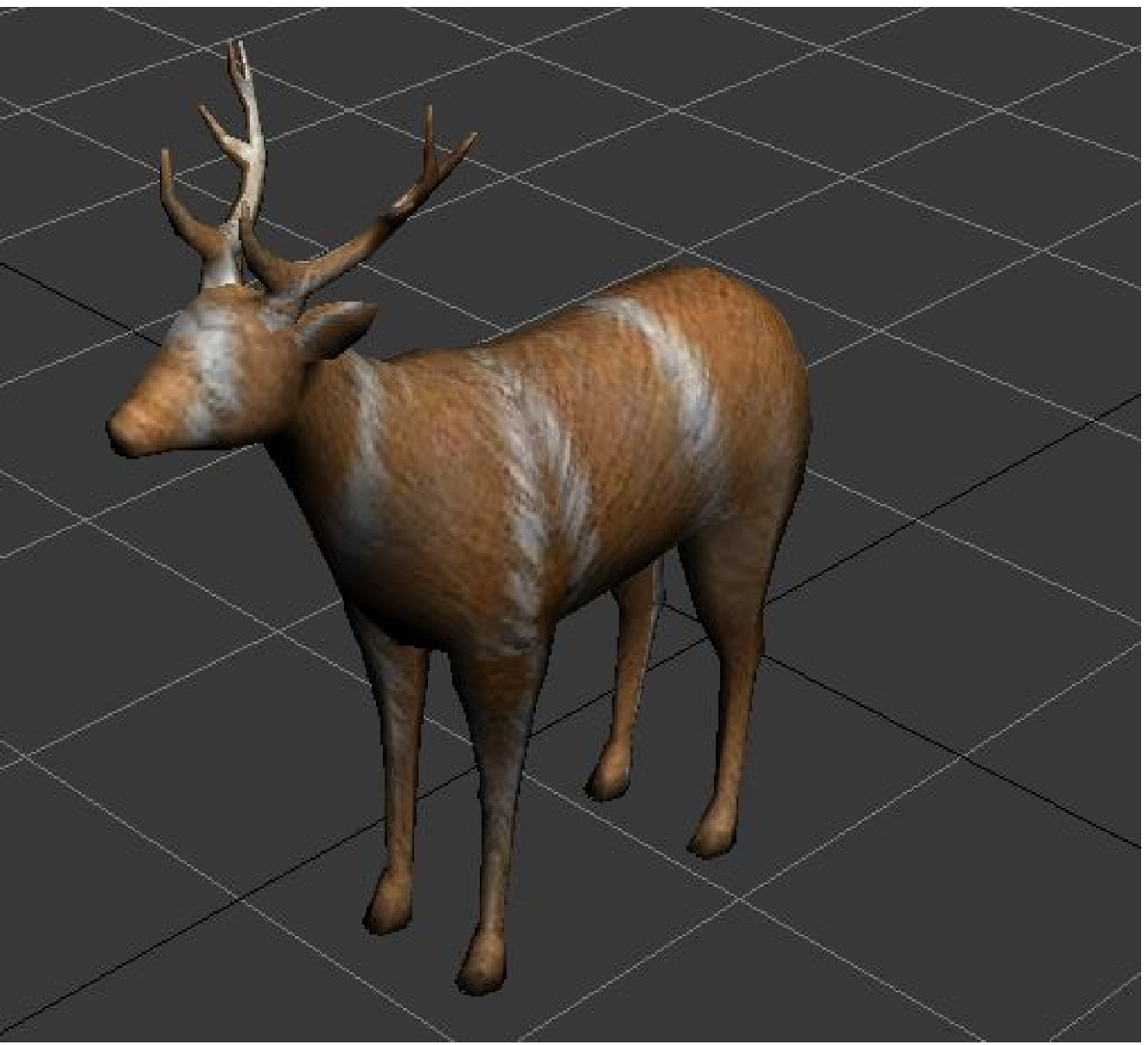}}	
&  693 &  470 &  692 & 1382 &  5 & 124 \\
\addlinespace[0.1cm]
\raisebox{-.5\height}{\includegraphics[width=0.15\textwidth,height=15mm]{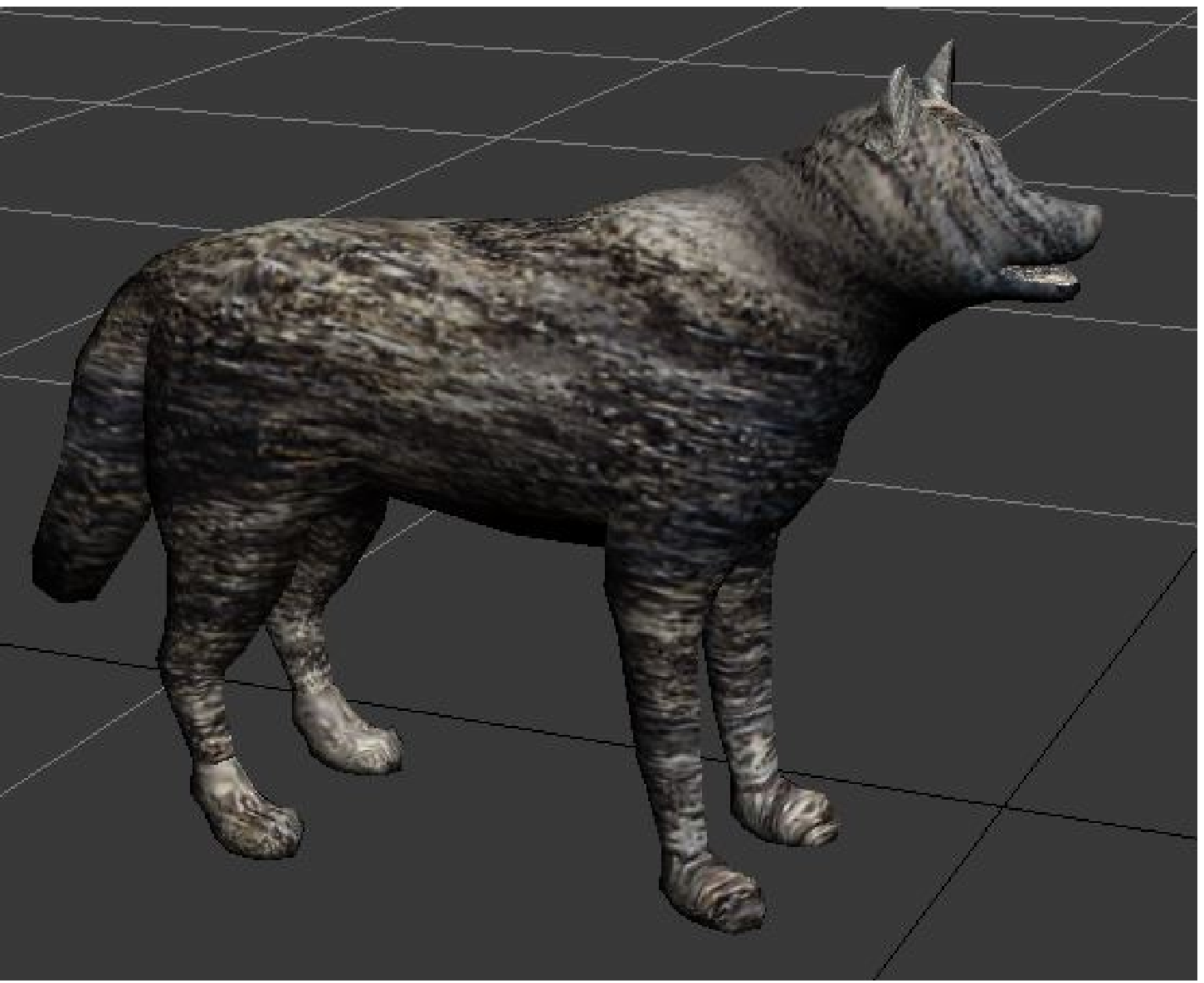}}		&  645 &  414 &  645 & 1286 &  5 & 113 \\
\addlinespace[0.05cm]
\bottomrule
\end{tabular*}
\end{table*}

\subsection{Visual Results}

\begin{figure}[!ht]
\centering
\subfloat{\includegraphics[width=0.22\linewidth,height=20mm]{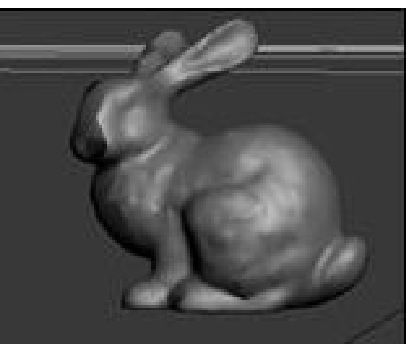}\label{fig:quad1}}\mbox{ }
\subfloat{\includegraphics[width=0.22\linewidth,height=20mm]{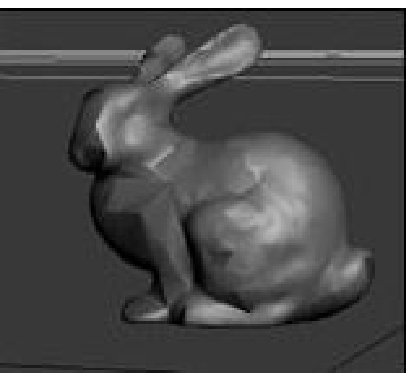}\label{fig:quad2}}\mbox{ }
\subfloat{\includegraphics[width=0.22\linewidth,height=20mm]{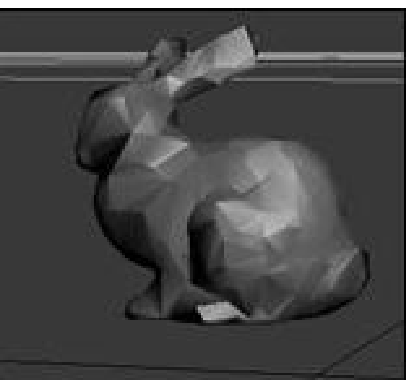}\label{fig:quad3}}\mbox{ }
\subfloat{\includegraphics[width=0.22\linewidth,height=20mm]{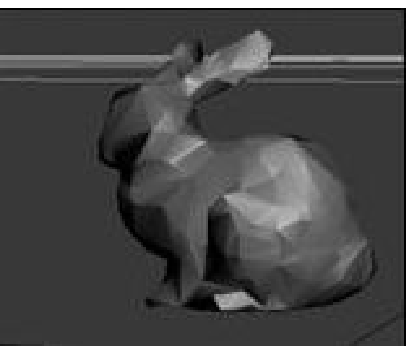}\label{fig:quad4}}\\\vspace{-5pt}
\subfloat{\includegraphics[width=0.22\linewidth,height=20mm]{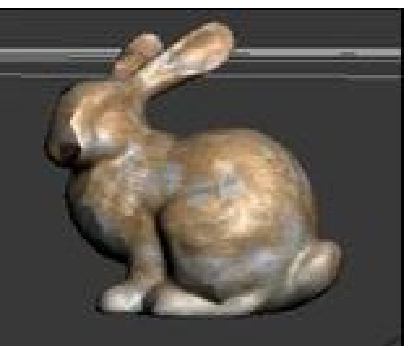}\label{fig:ias1}}\mbox{ }
\subfloat{\includegraphics[width=0.22\linewidth,height=20mm]{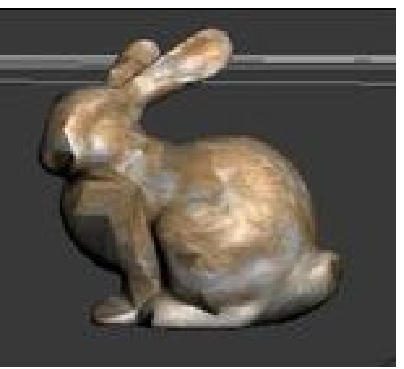}\label{fig:ias2}}\mbox{ }
\subfloat{\includegraphics[width=0.22\linewidth,height=20mm]{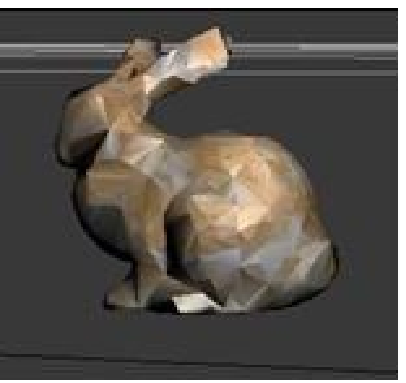}\label{fig:ias3}}\mbox{ }
\subfloat{\includegraphics[width=0.22\linewidth,height=20mm]{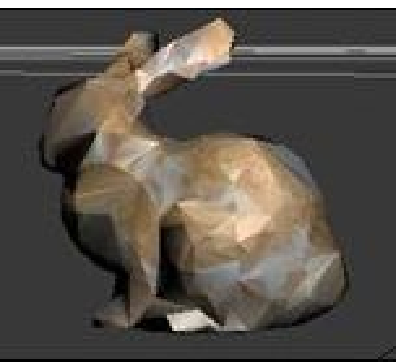}\label{fig:ias4}}\\\vspace{-5pt}
\subfloat{\includegraphics[width=0.22\linewidth,height=20mm]{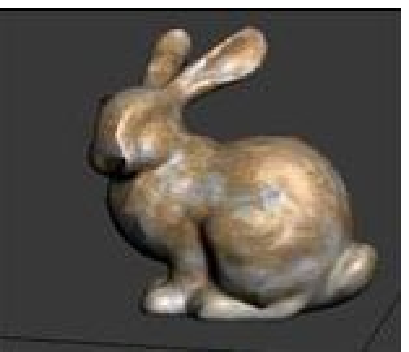}\label{fig:gaps1}}\mbox{ }
\subfloat{\includegraphics[width=0.22\linewidth,height=20mm]{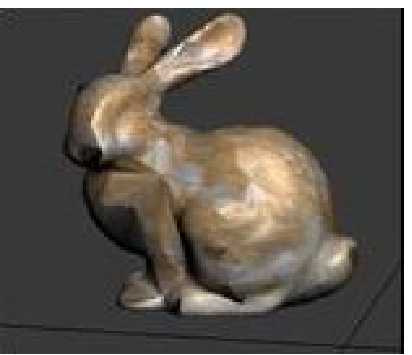}\label{fig:gaps2}}\mbox{ }
\subfloat{\includegraphics[width=0.22\linewidth,height=20mm]{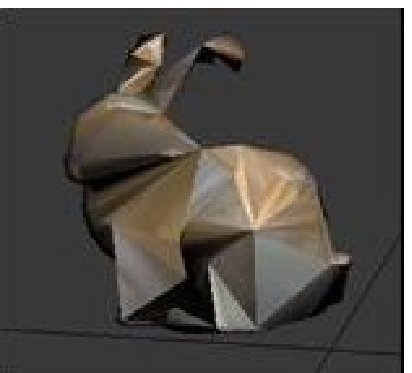}\label{fig:gaps3}}\mbox{ }
\subfloat{\includegraphics[width=0.22\linewidth,height=20mm]{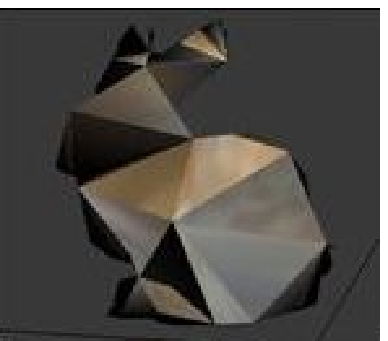}\label{fig:gaps4}}\\
\caption{Visual comparison of three different approaches on the Stanford bunny, Top row: Quadric simplification; Middle row: proposed ITS; and Bottom row: GAPS. Moreover, column-wise corresponds to different levels of simplification 10\%, 20\%, 50\% and 70\% respectively.}
\label{fig:cmp1}
\end{figure}

A comparison between the three different approaches to simplifying the Stanford bunny input model by reducing vertices and face counts is shown in Figure~\ref{fig:cmp1}. The simplification is done by quadric algorithm, GAPS, and proposed ITS at 10\%, 20\%, 50\% and 70\% level of simplification. The quadric approach does not simplify textures and normal coordinates; hence we have only accounted for geometry.

\begin{figure}[!ht]
\centering
\subfloat{\includegraphics[width=0.45\linewidth,height=35mm]{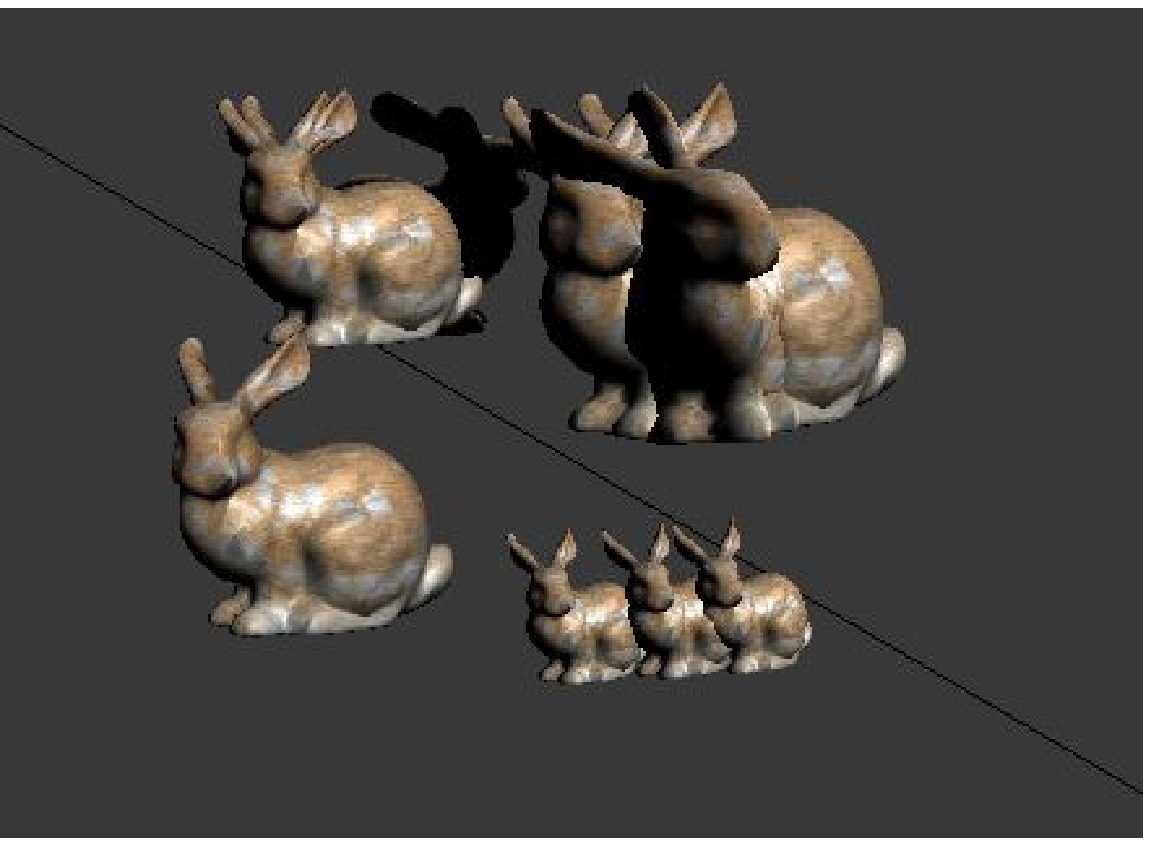}\label{fig:tb10}}\mbox{ }
\subfloat{\includegraphics[width=0.45\linewidth,height=35mm]{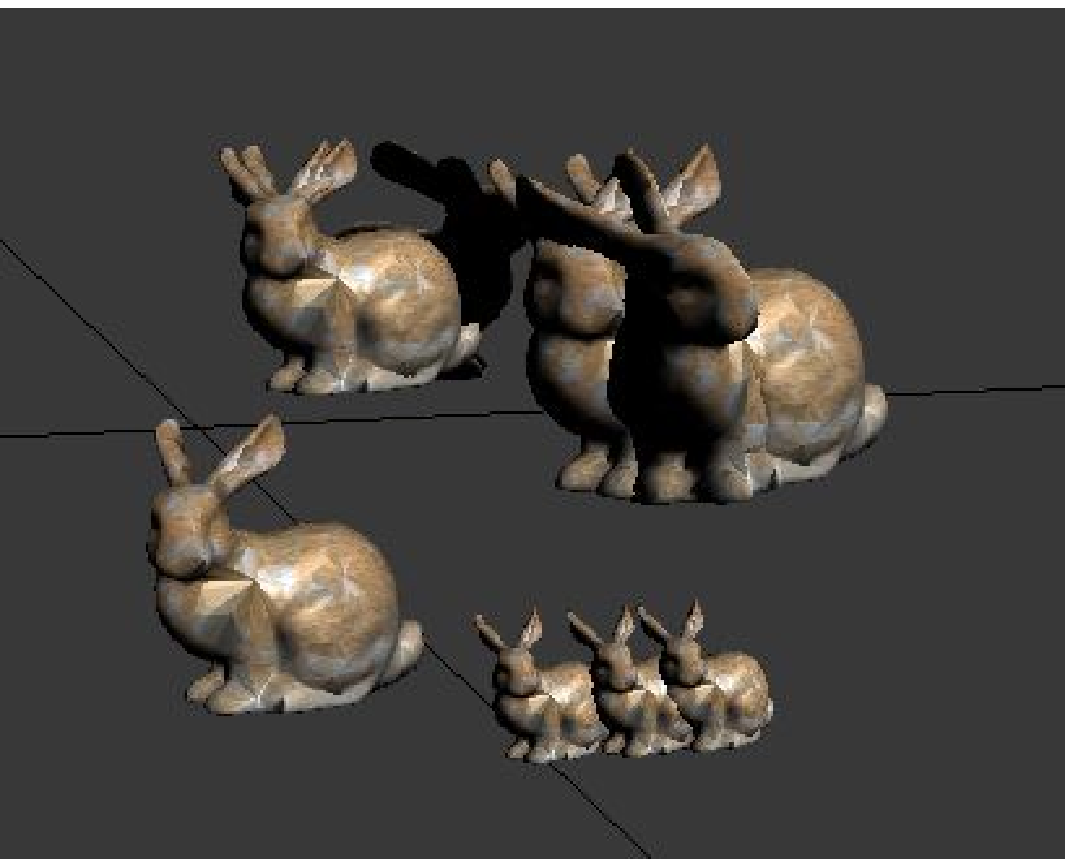}\label{fig:tb20}}
\caption{Results of simplification for 13 different instanced Stanford bunny, Top: with 10\%t fewer vertices and triangles count in 1.72 seconds; and Bottom: with 20\% fewer vertices and triangles count in 2.83 seconds.}
\label{fig:twelvebunny}
\end{figure}

Once quadric and attribute simplification is done, instancing is introduced. Then, simplified instanced 3D mesh copies are written as indexed triangles into an object file. Figure~\ref{fig:twelvebunny} shows twelve copies of 10\% and 20\% reduced vertices and face counts of Stanford bunny with different rotation angles, scales, and positions but the same geometry, computed in 1.72 and 2.83 seconds respectively.

\begin{figure*}[!ht]
\centering
\subfloat{\includegraphics[width=0.3\linewidth,height=2.5cm]{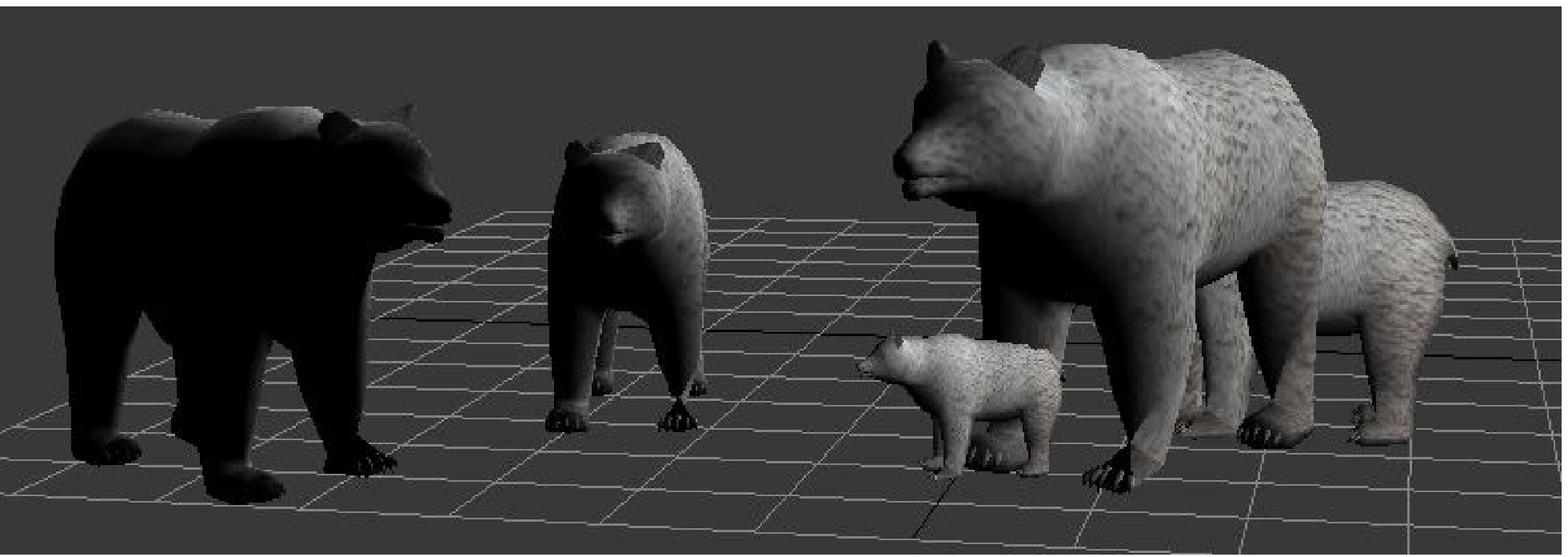}\label{fig:b10}}\mbox{ }
\subfloat{\includegraphics[width=0.3\linewidth,height=2.5cm]{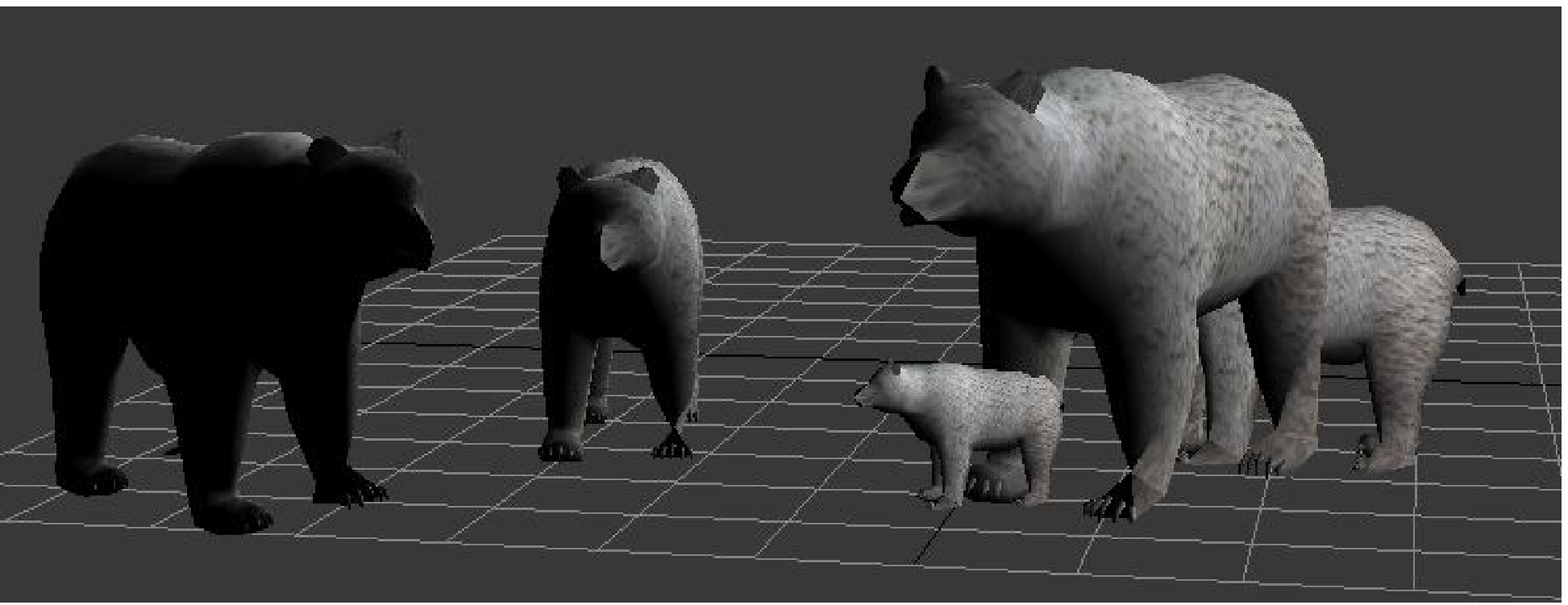}\label{fig:b20}}\mbox{ }
\subfloat{\includegraphics[width=0.3\linewidth,height=2.5cm]{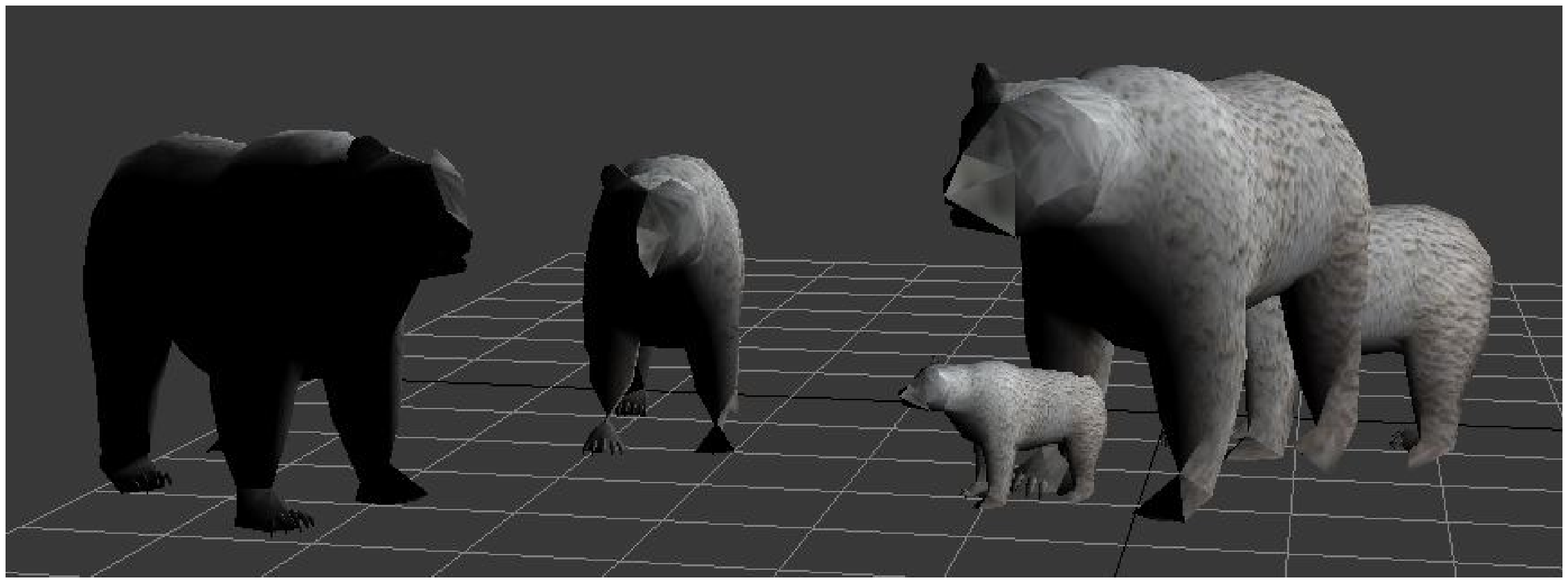}\label{fig:b50}}
\caption{Results of simplification for four different instanced bears, Left: with 10\% fewer vertices and faces count and four different instances produced in 0.78 seconds; Middle: with 20\% fewer vertices and faces count and four different instances produced in 0.97 seconds; and Right: with 50\% fewer vertices and faces count and four different instances produced in 1.07 seconds.}
\label{fig:bears}
\end{figure*}

\begin{figure*}[!ht]
\centering
\subfloat{\includegraphics[width=0.3\linewidth,height=2.5cm]{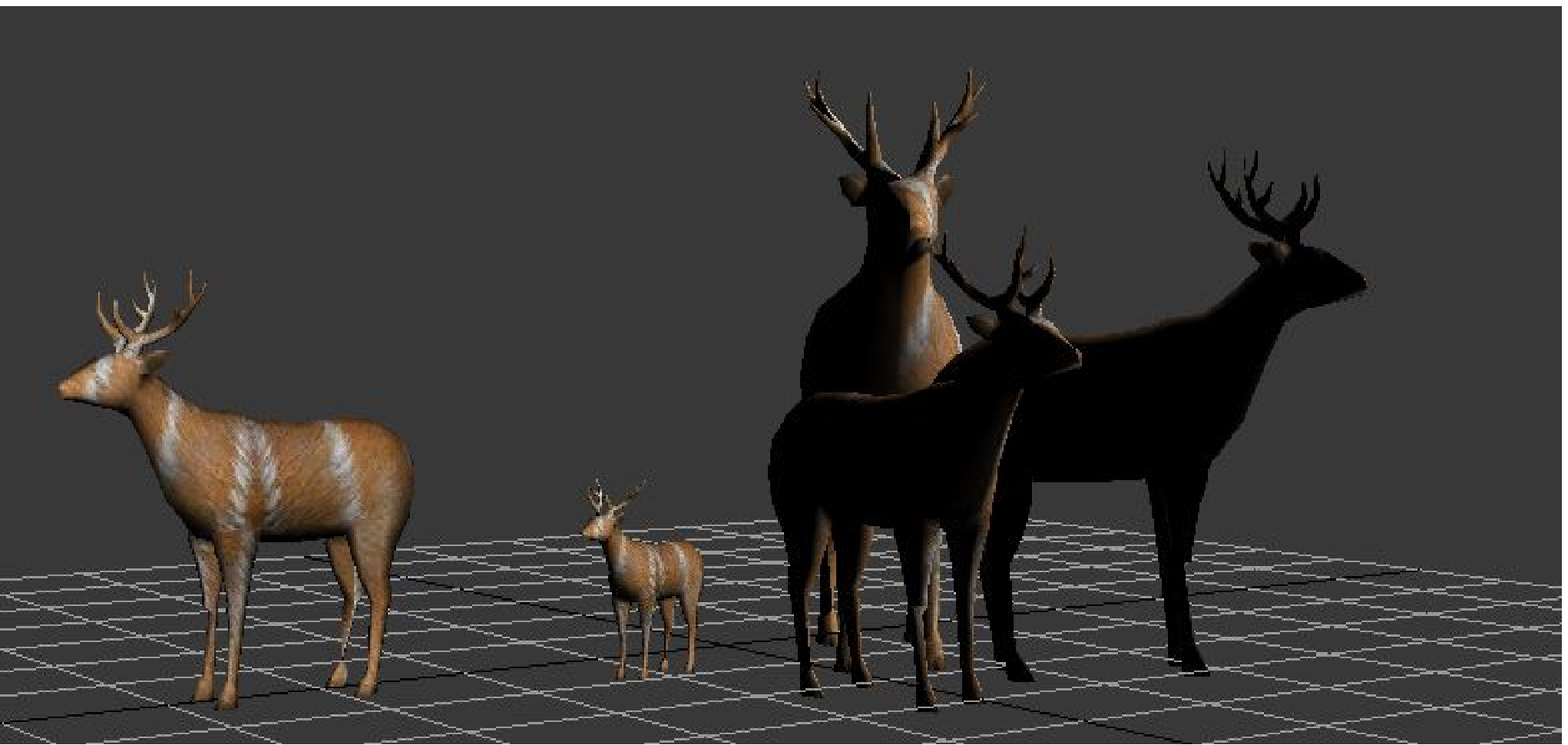}\label{fig:d10}}\mbox{ }
\subfloat{\includegraphics[width=0.3\linewidth,height=2.5cm]{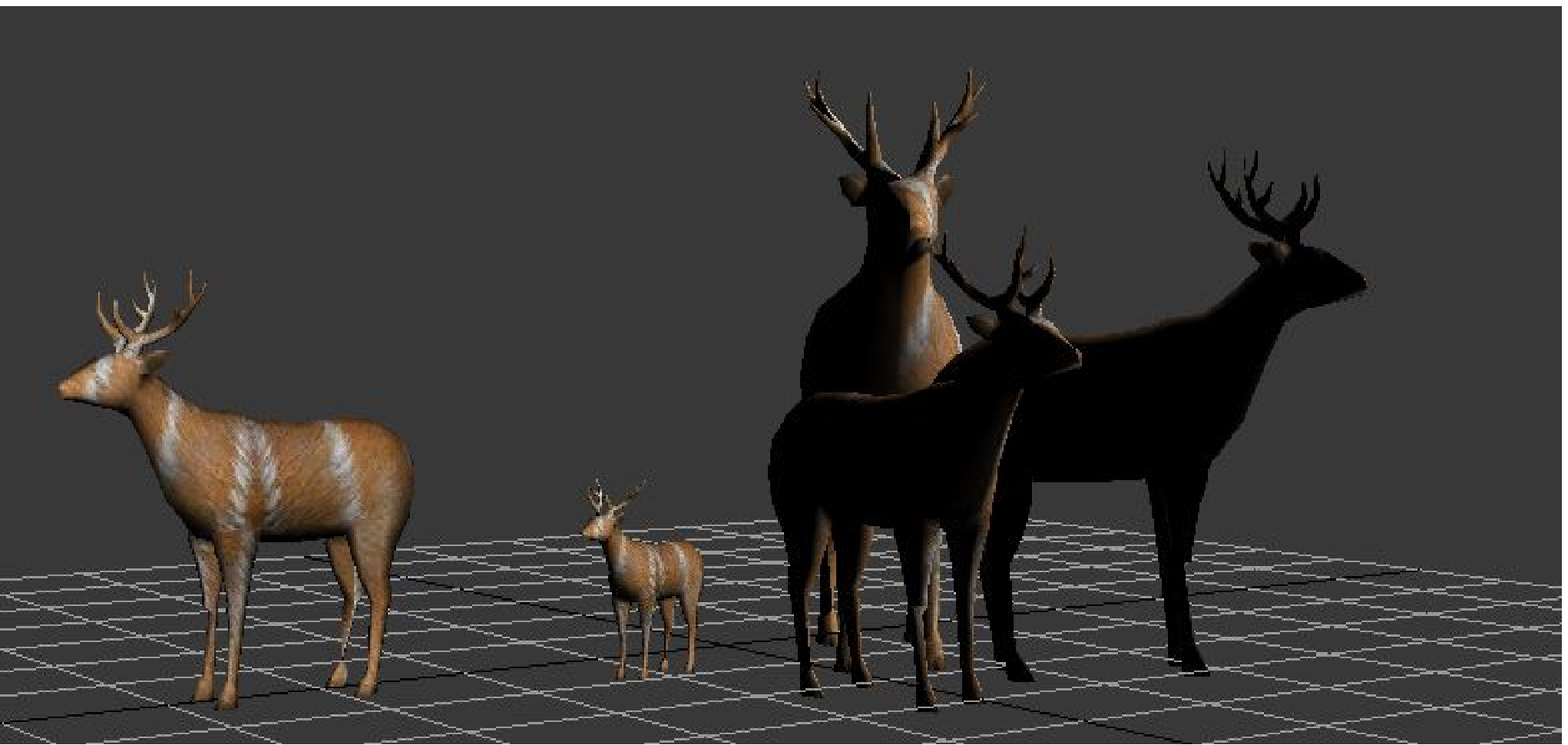}\label{fig:d20}}\mbox{ }
\subfloat{\includegraphics[width=0.3\linewidth,height=2.5cm]{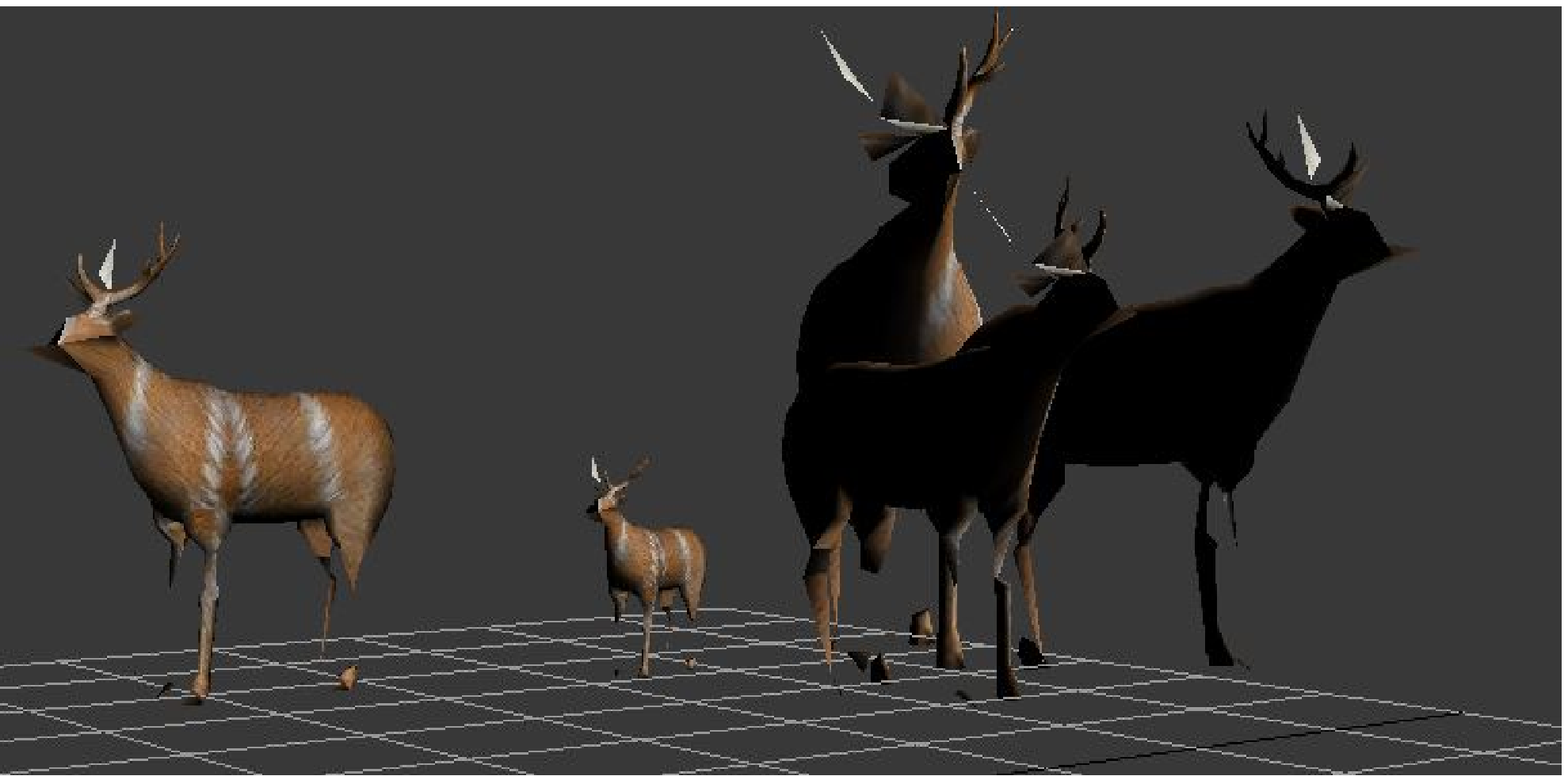}\label{fig:d50}}
\caption{Results of simplification for four different instanced deer, left: with 10\% fewer vertices and faces count and four different instances produced in 0.72 seconds; Middle: with 20\% fewer vertices and faces count and four different instances produced in 0.99 seconds; and Right: with 50\% fewer vertices and faces count and four different instances produced in 1.29 seconds.}
\label{fig:deer}
\end{figure*}

\begin{figure*}[!ht]
\centering
\subfloat{\includegraphics[width=0.3\linewidth,height=2.5cm]{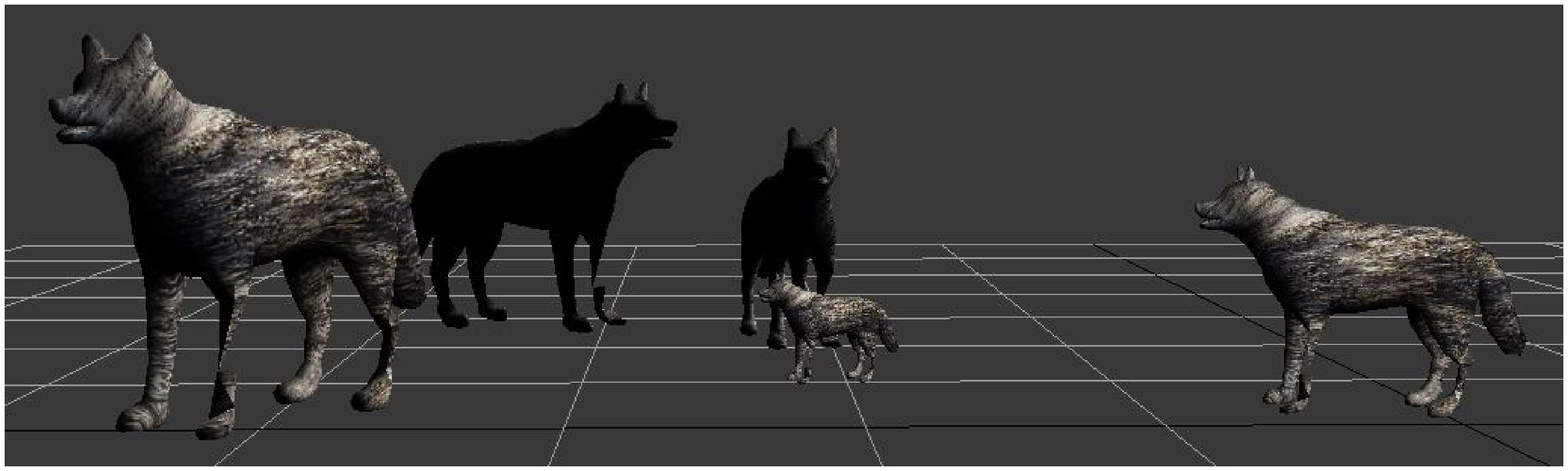}\label{fig:w10}}\mbox{ }
\subfloat{\includegraphics[width=0.3\linewidth,height=2.5cm]{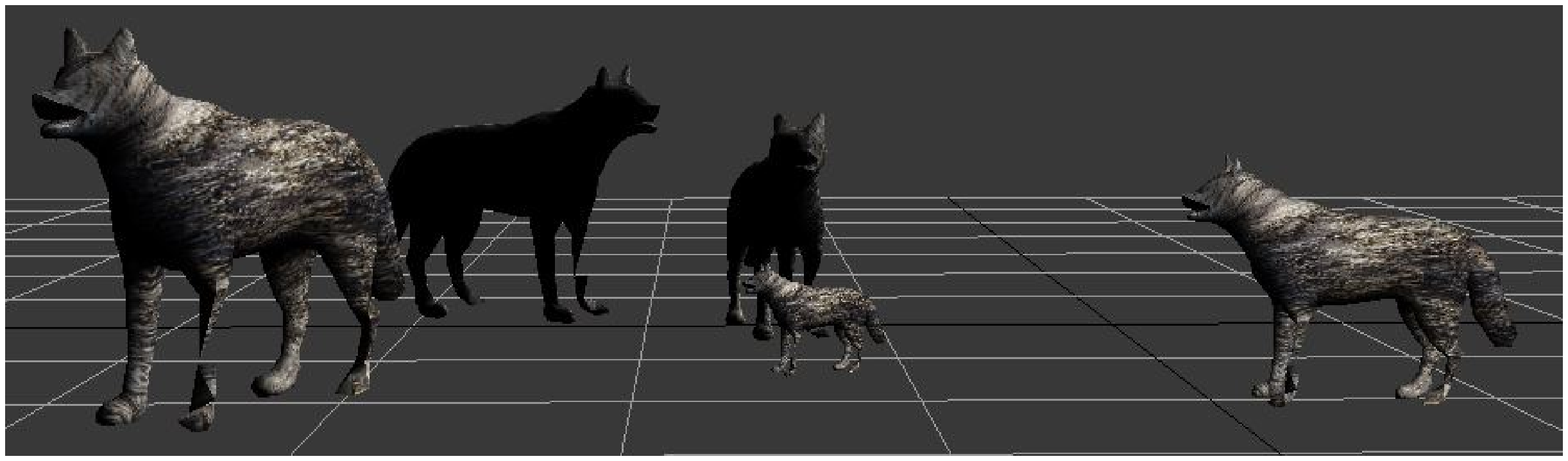}\label{fig:w20}}\mbox{ }
\subfloat{\includegraphics[width=0.3\linewidth,height=2.5cm]{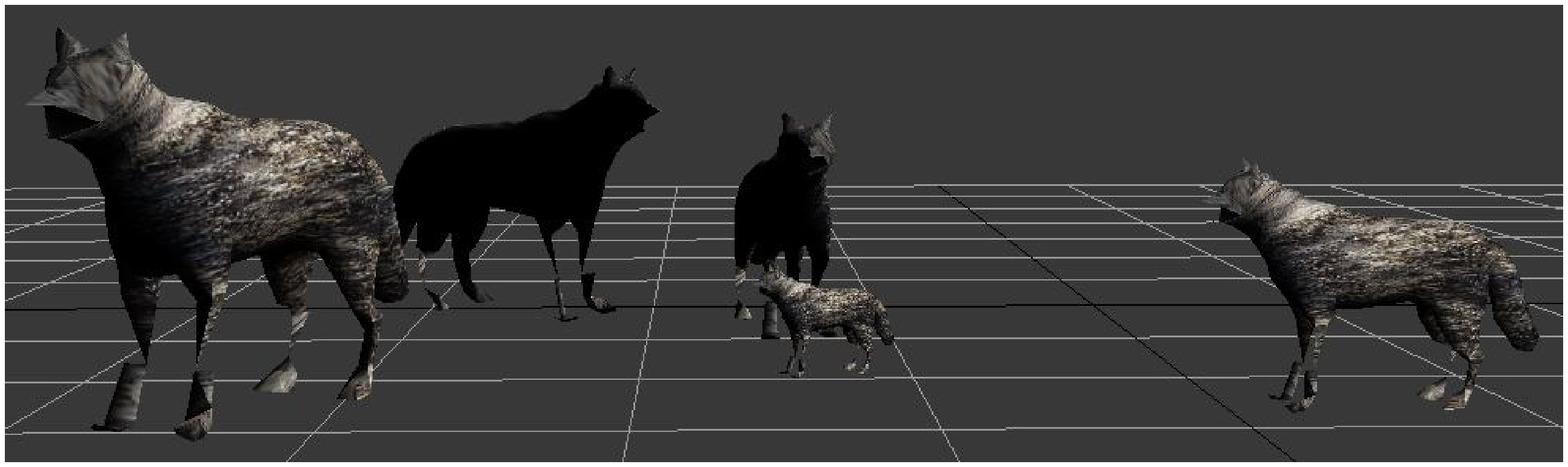}\label{fig:w50}}
\caption{Results of simplification for four different instanced Wolves, Left: with 10\% fewer vertices and faces count produced in 0.70 seconds; Middle: with 20\% fewer vertices and faces count produced in 0.84 seconds; and Right: with 30\% fewer vertices and faces count produced in 1.27 seconds.}
\label{fig:wolves}
\end{figure*}

Additionally, few other results of meshes simplification with instancing are presented. Figure~\ref{fig:bears} shows 10\%, 20\% and 50\% simplification of test input model of a bear. The model is simplified with four instances at different rotation angles, scales, and translations. It is evident from the result that after reducing 20\% of vertices, most of the model's shape is retained. The time taken by the proposed approach to reducing five models of the bear is 0.78, 0.97, and 1.07 seconds by 10\%, 20\%, and 50\% simplification respectively. Furthermore, Figure~\ref{fig:deer} shows 10\%, 20\% and 50\% simplification of test input model of a deer. The simplified model has four instances with different rotation angles, scales, and translations. The time taken by our system to reduce five models of a deer is 0.72, 0.99, and 1.29 seconds by 10\%, 20\%, and 50\% simplification respectively. Similarly, Figure~\ref{fig:wolves} shows four different instanced wolves simplified by a 10\%, 20\%, and 50\% simplification rate.

In the cases of deer, bears, and wolves, upon the reduction of face counts by 10\% and 20\%, most of the details and shapes were retained. For example, in the case of the deer, the shape of its mouth and sharpness of its antlers were still there. However, when reduced to 50\% face count, the face started losing its shape and so did the antlers. Its legs also disappeared partially and the tail disappeared completely. These results also reveal another fact about this approach: the results get better with the increase in data size. As Stanford bunny has vertices and faces in thousands, thus, upon reducing its face count to 50\%, loss of information was enough to be noticed from a distance.

\subsection{Result Statistics}

\begin{table*}[!ht]%
\centering
\caption{Faces and vertices count chart of simplified result models with original inputs (without instances).\label{tbl:res1}}
\begin{tabular*}{1.0\linewidth}{@{\extracolsep\fill}lcc|cc|cc|cccclD{.}{.}{3}l@{\extracolsep\fill}}
\toprule
Models & 
\multicolumn{2}{c|}{Original} & 
\multicolumn{2}{c|}{10\% Simplified} & 
\multicolumn{2}{c|}{20\% Simplified} & 
\multicolumn{2}{c}{50\% Simplified} \\
\cmidrule{2-9}
& Vertices & Faces & Vertices & Faces & Vertices & Faces & Vertices & Faces \\ [0.5ex]
\midrule
Bunny 	& 2503 & 4968 & 2247 & 4456 & 1998 & 3958 & 1244 & 2448 \\
Bear 	&  742 & 1360 &  664 & 1204 &  589 & 1054 &  370 &  652 \\
Deer 	&  693 & 1382 &  623 & 1234 &  550 & 1082 &  345 &  650 \\
Wolf 	&  645 & 1286 &  581 & 1156 &  516 & 1024 &  317 &  620 \\
\bottomrule
\end{tabular*}
\end{table*}

Table~\ref{tbl:res1} shows a reduction in faces and vertices in comparison with the original number of vertices and faces. Here, reduction in vertex counts by 10\%, 20\% and 50\% level of simplification are presented. The time taken to reduce input test models to the desired level using this approach is shown in Table~\ref{tbl:res3}. Moreover, the reduction in file size is also mentioned in this table.


A comparison between un-instanced (Quadrics and GAPS) and instanced (IAS and proposed ITS) is shown in Table ~\ref{tbl:res3} in terms of time taken and size of the output files produced. It should be noted that GAPS does not support instancing, thus in order to reduce instanced meshes, it took all the instances as an indexed triangle mesh, taking much more time than the proposed instanced version. Moreover, the file size produced by GAPS was also larger compared to the proposed approach.


\begin{table*}[!ht]%
\centering
\caption{Comparison between un-instanced (Quadrics and GAPS) and instanced (IAS and proposed ITS) in terms of time and space required. Time in seconds and size in kilobytes.\label{tbl:res3}} 
\begin{tabular*}{1.0\linewidth}{@{\extracolsep\fill}lccccc|cccc|ccccD{.}{.}{3}l@{\extracolsep\fill}}
\toprule
\multicolumn{2}{c}{~} 
& \multicolumn{4}{c|}{10\% Simplified}
& \multicolumn{4}{c|}{20\% Simplified}
& \multicolumn{4}{c}{50\% Simplified} \\
\multicolumn{2}{c}{~} 
& \multicolumn{2}{c}{Un-instanced}
& \multicolumn{2}{c|}{Instanced}
& \multicolumn{2}{c}{Un-instanced}
& \multicolumn{2}{c|}{Instanced}
& \multicolumn{2}{c}{Un-instanced}
& \multicolumn{2}{c}{Instanced} \\
\midrule
\multicolumn{2}{c}{~} & Q	& GAPS	& IAS	& ITS	& Q	& GAPS	& IAS	& ITS	& Q	& GAPS	& IAS	& ITS \\
\midrule
\multirow{2}{*}{Bunny}	& Size (KB)	& 6383	& 6370	& 4672	& 4661	& 5928	& 5915	& 4147	& 4142	& 4537	& 4537	& 2594	& 2576 \\
							& Time	(ms) 	& 10140	& 10517	& 741	& 774	& 12922	& 12662	& 910	& 977	& 8125	& 8151	& 602	& 621 \\
\midrule
\multirow{2}{*}{Bear}		& Size	(KB) & 565	& 565	& 470	& 469	& 515	& 515	& 417	& 415	& 385	& 385	& 268	& 269 \\
							& Time	(ms)	& 945	& 895	& 193	& 185	& 1130	& 1115	& 223	& 225	& 1690	& 1565	& 321	& 287 \\
\midrule
\multirow{2}{*}{Deer}		& Size	(KB) & 550	& 550	& 467	& 465	& 505	& 505	& 414	& 414	& 375	& 375	& 265	& 264 \\
							& Time	(ms)	& 975	& 2250	& 199	& 215	& 1175	& 1195	& 228	& 238	& 1665	& 1820	& 341	& 406 \\
\midrule
\multirow{2}{*}{Wolf}		& Size	(KB)	& 505	& 505	& 427	& 430	& 465	& 465	& 383	& 384	& 345	& 345	& 248	& 244 \\
							& Time	(ms)	& 835	& 1155	& 192	& 174	& 945	& 1120	& 203	& 198	& 1445	& 1475	& 328	& 322 \\
\bottomrule
\end{tabular*}
\end{table*}

\begin{figure}[!ht]
\centering
\subfloat{
\includegraphics[width=1.0\linewidth]{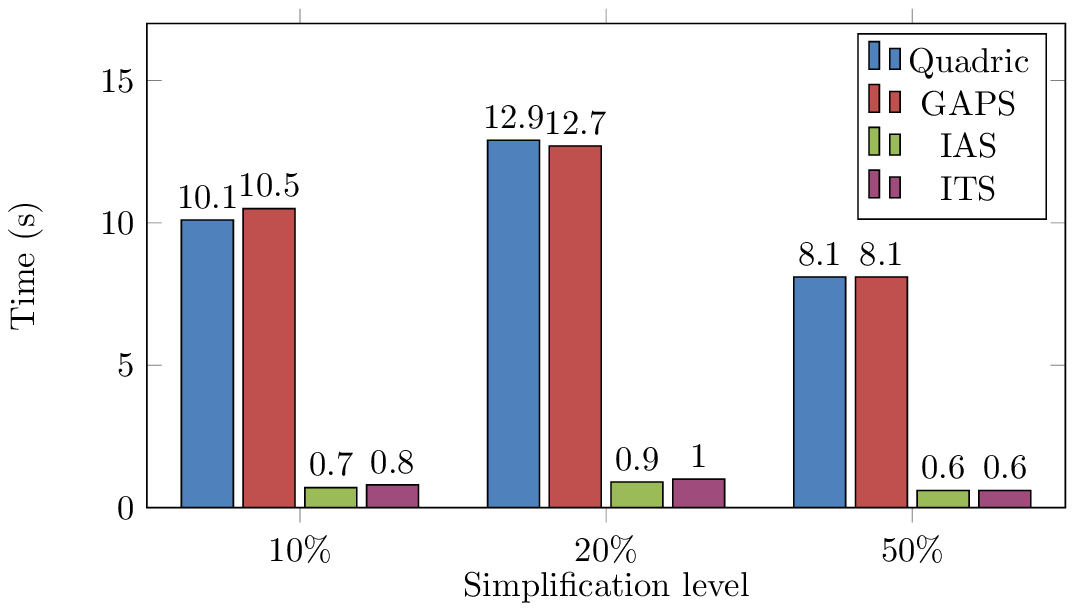}
\label{fig:cmptime}}

\subfloat{
\includegraphics[width=1.0\linewidth]{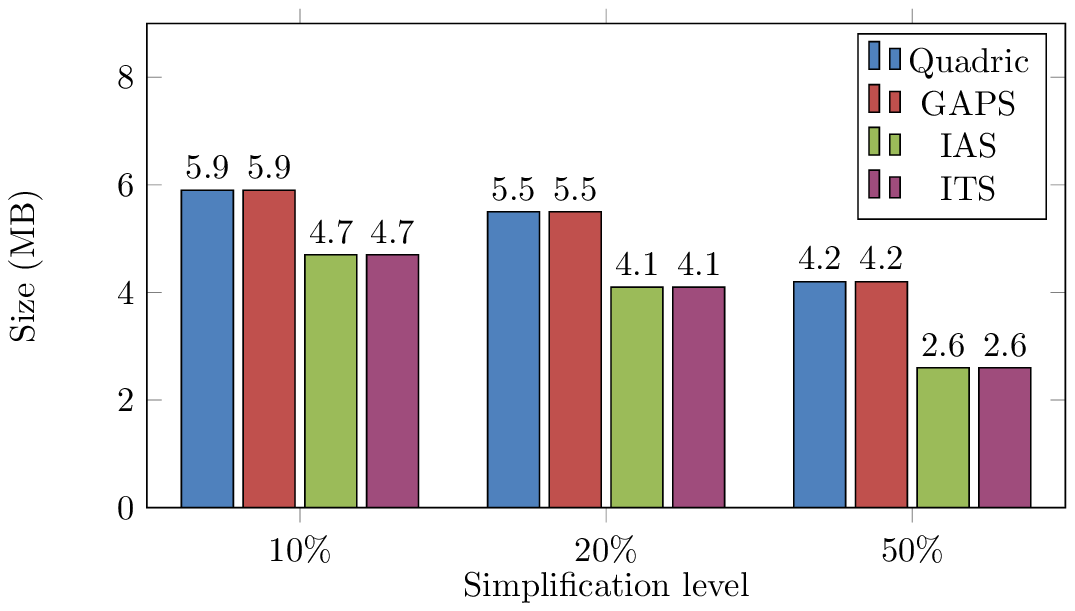}
\label{fig:cmpsize}}
\caption{Time and size comparison graphs of Instanced Stanford Bunny and Un-Instanced Stanford Bunny, simplified by 10\%, 20\%, and 50\%.}
\label{fig:cmptmp}
\end{figure}

Figure~\ref{fig:cmptmp} shows a comparison of processing times and the result file sizes for simplification of Stanford bunny using un-instanced (Quadrics and GAPS) and instanced (IAS and ITS).

\begin{figure}
\centering
\subfloat{
\includegraphics[width=1.0\linewidth]{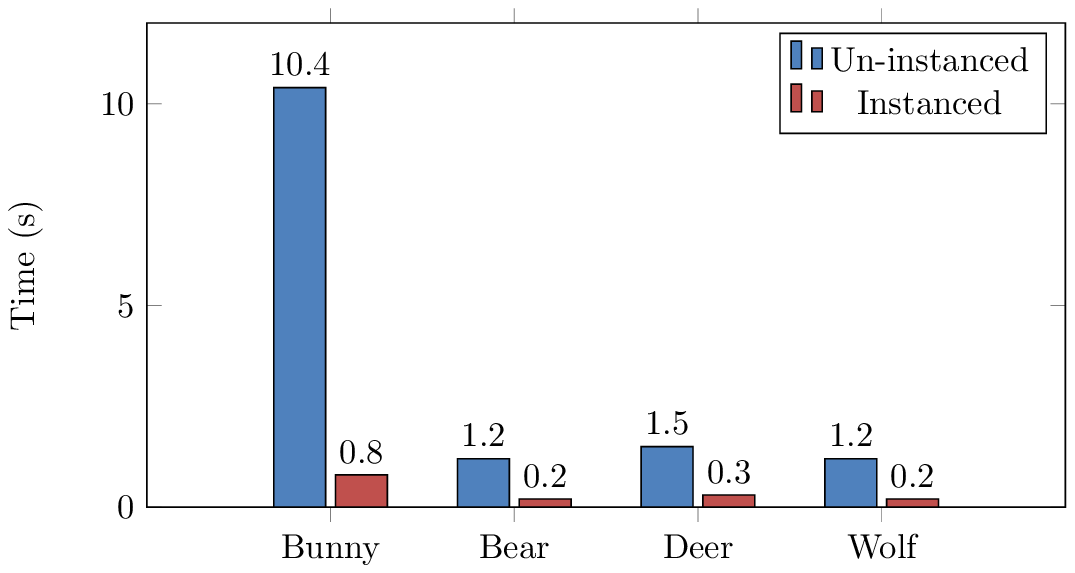}
\label{fig:avgtime}}

\subfloat{
\includegraphics[width=1.0\linewidth]{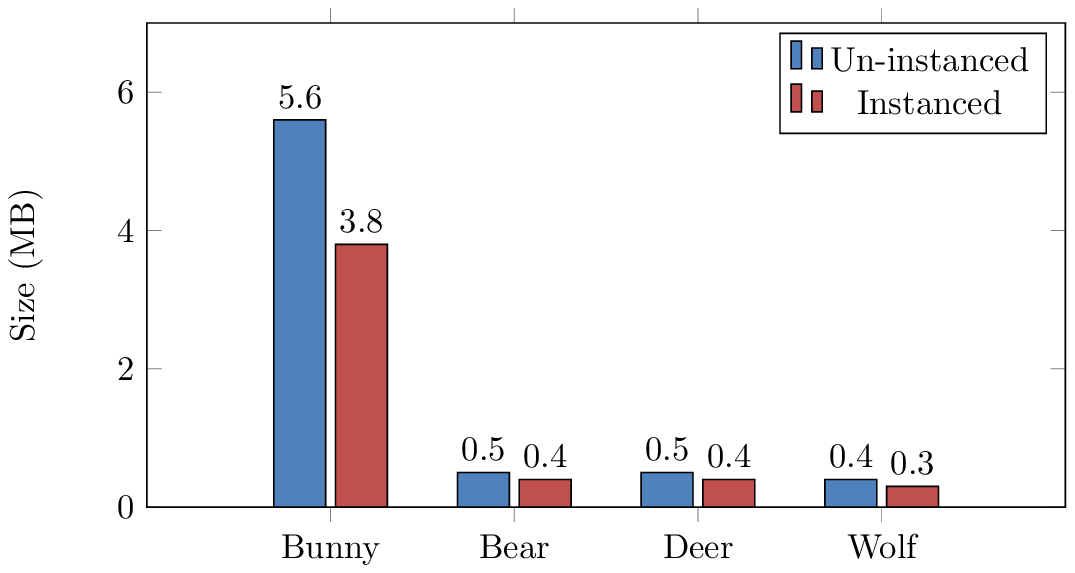}
\label{fig:avgsize}}
\caption{Average time and size for Instanced and Un-Instanced simplication at levels 10\%, 20\%, and 50\%.}
\label{fig:avgtimesize}
\end{figure}

Time and size comparison for un-instanced (Quadrics and GAPS) and instanced (IAS and proposed ITS) using different input 3D models: Stanford bunny, bear, deer, and wolf at different levels of simplification are shown in Figure~\ref{fig:avgtimesize}.

The proposed simplification method produces results significantly more efficient than previous approaches. The results clearly show its efficiency compared to GAPS. Using this approach, 3D content with textures and normals can be shared using minimal network bandwidth; thereby contributing to making 3D content readily available, in particular for virtual worlds. Moreover, there is a significant processing time reduction while producing acceptable results. All in all, the results make this work a good contribution in simplification of 3D textured instance aware models.

\section{Discussion}
\label{sec:discussion}


We proposed a hybrid 3D textured model simplification approach that is faster and efficient. This approach is a combination of three variant ideas to deal with three objectives: it simplifies geometry, it simplifies vertex attributes, and it controls file sizes by accounting for instances. There are many approaches that simplify geometry in a very efficient way, whereas other approaches deal with attributes, however, they lack in the reduction of file size. Our algorithm is efficient enough to handle geometry and attribute simplification while resulting in smallest possible file size. 

Furthermore, the proposed simplification was split into two stages: geometric simplification followed by attribute simplification, to get a better geometric appearance. This was mainly due to the assumption that the visual quality of a 3D model is more affected by its geometry rather than its textures or normals. Thence, a geometric error was used to make an initial vertex collapse. In case the collapse kept the geometric quality within acceptable limits, it was further simplified using vertex attributes.

The proposed method is more generalized, as most of the 3D content is stored as an instanced file containing textures and normal coordinates. It is noteworthy that most of the virtual environments and 3D worlds use instanced data, which was not the case in previous approaches. For example, instanced simplification uses instancing but lacks accounting for attributes while GAPS does not use instanced input, rather it takes them as indexed triangles for applying quadric simplification.

We came across a few challenges in this work: First, instanced 3D mesh input file formats like COLLADA, Autodesk Maya, and Autodesk 3ds Max files are not readable in raw, making it almost impossible to take a deep look into the processing applied to 3D data. Object file format is the simplest readable format to carry such data, but a major issue with object file format is that it does not support instancing. Therefore, an input file format that is readable yet supports instancing needs to be developed. To overcome this challenge, we developed an instance supporting version of object file format. Second, to write the instances supporting object file, we developed the transformation matrices because this information is given by default in files like COLLADA and Autodesk Maya. Third, making a new file format raised issues such as which visual tool would open a file type. To resolve this issue, we came up with the solution that data is to be read from a customized object file and stored separately. This procedure was carried out on the data and the result was written as an indexed triangle mesh for visual analysis. Last, we also developed two different formats for the result file: one with indexed triangles, and the other with instances and transformations. The output produced using the proposed approach was viewable in different tools irrespective of instancing. Once the input 3D model was simplified with its geometry and other properties, the resulting simplified data was applied across all respective instances to get the desired result.

Some points require improvements, such as threshold selection is iterative and inefficient. Furthermore, the approach used to select valid pairs for collapse can be replaced with more sophisticated methods; this most likely can help generate better results. Point cloud idea is a very approximate and average method to calculate attribute errors that should definitely be improved. Geometry simplification can also be improved by finding optimal point for the collapse of a pair to make results more defined and qualitative.

The proposed solution restrictively takes triangulated input, it is not designed to deal with other polygons. Moreover, it is not tested in a complete 3D environment or some other real-world data, rather the evaluation is carried out using laboratory produced instanced meshes. Another constraint is the use of customized object type input as traditional object file format which does not support instances. Lastly, we assume that there is one mesh with multiple instances; however, this can be generalized for multiple meshes with many instances.

\section{Conclusion}
\label{sec:conclusion}


The proposed 3D simplification algorithm takes as input a 3D instanced model with attributes like geometry, normals, and textures. It uses iterative pair contractions based on surface attributes to simplify the models and uses quadric error matrices to find errors as the models get simplified. The resulting mesh simplification is used to simplify and produce multiple simplified meshes (instances). On larger 3D input models, the algorithm produces far better results compared to other approaches. Moreover, the time to reduce instanced meshes is equivalent to one mesh simplification. In the end, the results produced for 3D textured models are as refined as those produced using GAPS with efficient time-space and visual quality.

\bibliographystyle{IEEEtran}
\bibliography{egbib}

%
%
%
%

\end{document}